\documentstyle[nato,citesort]{crckapb}

\begin{opening}

\title{Dynamic Properties of Stretched Water}

\author{P. A. Netz$^1$}

\author{F. W. Starr$^2$}

\author{ H. E. Stanley$^3$}

\author{M. C. Barbosa$^{3,4}$} 

\institute{
$^1$ Departamento de Qu\'{\i}mica, Universidade Luterana do
Brasil,  92420-280, Canoas , RS, Brazil \\
$^2$ Polymers Division and Center for Theoretical and
Computational Materials Science, National Institute of Standards and
Technology, Gaithersburg, MD 20899, USA \\
$^3$ Center for Polymer Studies and Department of Physics,
Boston University, Boston, MA 02215, USA \\     
$^4$ Instituto de F\'{\i}sica, Universidade Federal do Rio
Grande do Sul, Caixa Postal 15051, 91501-970, Porto Alegre,
RS, Brazil}

\end{opening}

\begin{document}

\begin{abstract}

We investigate the dynamics of the extended
simple point charge (SPC/E) model of  water in the
supercooled region. The dynamics at negative
pressures show a minimum in the diffusion 
constant $D$ when the density is decreased at constant temperature,
complementary to the known maximum of $D$ at higher pressures.
A similar trend in the rotational diffusion is also observed.
\end{abstract}

\section{Introduction}

The thermodynamic description of supercooled water has
been a major topic of research  already for many years.
Most of this scientific effort has been concentrated in 
understanding the static anomalies present in   this complex fluid.
 It expands on freezing and,
at a pressure of 1 atm, the density has a maximum at
4$^\circ$C. Additionally, there is a minimum of the isothermal
compressibility at 46$^\circ$C and a minimum of the isobaric heat
capacity at 35$^\circ$C \cite{dou98}. These anomalies are linked with
the microscopic structure of liquid water, which can be regarded as a
transient gel---a highly associated liquid with strongly directional
hydrogen bonds \cite{geig79,Stanley80}. Each water molecule acts as both
a donor and an acceptor of bonds, generating a structure that is locally
ordered, similar to that of ice, but maintaining the long-range disorder
typical of liquids.

Several scenarios have been proposed to explain these anomalies. In
the `` stability limit conjecture '' \cite{spe82b,spe87}, the liquid 
spinodal line for water
is reentrant. It has a minimum at negative pressures and passes back to
positive $P$ as the temperature decreases. The increasingly anomalous
thermodynamic behavior  of liquid water as it is  cooled  at positive
pressures can be interpreted in terms of approaching this spinodal.
The `` critical point  hypothesis '' proposes a second critical
point as the terminus of a phase coexistence between a high density
and a low density liquids. The increase in the response functions 
is then interpreted as a signature of being in the vicinity of 
this critical region. The `` singularity-free hypothesis '' 
suggests that there is no divergence in the response functions. They
grow but stay finite \cite{Stanley80,sas96,sas98}.

Recently, dynamic properties of water have gained  attention 
both experimentally \cite{jon76,pri87} and in  computer
simulations \cite{ram87,sci91,bae94,har97,francesco,sta99e,scala}.
The surprising result of these works is that water also
exhibits an anomalous dynamical behavior. 
The increase of the applied pressure leads to an increase in water
translational diffusion coefficient and to a faster 
rotational diffusion~\cite{bag97,jon76,pri87}. These effects
can be understood as follows. The increase of pressure leads to 
an increase in  the number of defects  
and  in the presence of interstitial water. This disrupts the 
tetrahedral local structure, weakening the hydrogen bonds, and thus
increasing the diffusion constant \cite{sta99e,scala}. 
A further increase in the pressure leads to steric effects which works in
the direction of lowering the mobility.
The interplay of these factors
leads to a maximum in the diffusion constant \cite{sta99e,scala} at some
high density $\rho_{\mbox{\scriptsize max}}$.
As a result,for each isotherm a maximum of the diffusion coefficient 
and a  minimum in the
rotational correlation time are found.
However, the behavior at very low $\rho$ is less well understood.

 In this manuscript, we present our recent studies of how  the dynamics of
low-temperature water are affected by the decrease of the density \cite{Ne01}. We
perform molecular dynamics (MD) simulations of the SPC/E model of water
in the range 210~K $< T <$ 280~K and 0.825~g/cm$^3 < \rho <$
0.95~g/cm$^3$. We  calculate the rotational and 
 translational diffusion coefficients in this region. A relationship
between the behavior of the two coefficients is  suggested.

\section{Results}

We performed molecular dynamics simulations using 216 water molecules
described by the extended simple
point charge (SPC/E) model~\cite{ber87b}, in the canonical
ensemble (NVT), in a cubic simulation box using periodic
boundary conditions.
The diffusion coefficient $D$ was calculated using the slope of the linear
regression of the mean square displacement versus time using a range
of time long enough to assure that the molecules have reached a diffusional
behavior.  We show $D$ along
isotherms in Figure $1$.  For $T\le260$~K, $D$ has a minimum value at
$\rho\approx 0.9$~g/cm$^3$, which becomes more pronounced at lower $T$
(Figure $1$).  This behavior can be understood considering the
structural changes that occur with decreasing density. At low $T$, the
decreased density enhances the local tetrahedral ordering, which leads
to a decrease in $D$.  Further decreases in density reduces the
stability of the tetrahedral structure and causes an increase of $D$.
The location of the minimum is near the ice Ih density $\approx
0.915$~g/cm$^3$, which is the density where the perfect tetrahedral
order occurs.
The orientational relaxation was analyzed using 
the rotational autocorrelation functions~\cite{all87,yeh99}:

$$
C^{(i)}({\bf e}) = \langle P_i [{\bf e} (t) \cdot {\bf e} (0)]
                  \rangle
$$ 

where ${\bf e}$ is a chosen unity vector describing the orientation of
the molecules and $P_i$ is the $i$-th order Legendre Polynomial (we restrict
ourselves only to the study of the first two correlations):

$$
P_1 (x) = x
$$

$$
P_2 (x) = \frac{3}{2} x^2 - \frac{1}{2}
$$

We choose three vectors to describe the orientation.
The  first vector is the unit vector with the same orientation
as the dipole moment. The second corresponds to the O-H bond direction and
the third is a vector perpendicular to the plane of the molecule.\\
The correlation functions were fitted to a biexponential decay 
function~\cite{yeh99}, and two relaxation times were calculated,
corresponding to a slower (I) and a faster (II) mode.

$$
C = a_0 \exp (-b t^2/2) + a_I \exp (-t/\tau^{I}) + a_{II} \exp (-t/\tau^{II})
$$

For each correlation using a given vector, two times were
calculated, leading to a rather complicated symbology. 
For example, the slower
relaxation time for the second-order correlation using
the third vector would be $\tau^{I,(2)}_3$.
With few exceptions, however, we have found only one relaxation time, 
for correlation in a given vector, 
that is $\tau^{I,(x)}_y$ = $\tau^{II,(x)}_y$. 
We obtain therefore six correlation times for each simulated point, 
$\tau^{(1)}_1$ and  $\tau^{(2)}_1$ for the first orientation vector and 
similarly $\tau^{(1)}_2$, $\tau^{(2)}_2$,
$\tau^{(1)}_3$ and  $\tau^{(2)}_3$.
At T = 240 K where a detailed analysis was
carried out, we find a maximum in the
orientational time as illustrated in Figure $2$.

\section{Conclusions}

We analyze the dynamic properties of
supercooled water.
  For high densities
($\rho>\rho_{\mbox{\scriptsize max}}$), water behaves as a normal liquid
and the decrease of translational diffusion coefficient, $D$, with increasing pressure is governed by steric
effects. For $\rho_{\mbox{\scriptsize min}} < \rho <
\rho_{\mbox{\scriptsize max}}$, as the pressure is decreased, the
presence of defects and interstitial water decrease, the tetrahedral
structure dominates, with stronger hydrogen bonds. This process reaches
its maximum at $\rho=\rho_{\mbox{\scriptsize min}} \approx
\rho_{\mbox{\scriptsize ice}}$. Further stretching destabilizes the
hydrogen bond network, leading to an increase in mobility. 
Preliminary studies of the rotational diffusion show
a maximum in the orientational time
at low density region in compass with the behavior of $D$.

\section{Acknowledgments}

We thank the National Science Foundation (NSF), the Conselho Nacional de
Desenvolvimento Cientifico e Technologico (CNPq), the Fundacao de Amparo
a Pesquisa do Rio Grande do Sul (Fapergs) for financial support.  FWS
thanks the National Research Council for financial support.

\section{Caption Figures}

Figure 1:(a) Dependence of the diffusion constant $D$ on $\rho$ along
  isotherms (for $\rho \le 1.0$~g/cm$^3$).  Open symbols are 
from ref. ~ \cite{Ne01}, and filled symbols are from ref.~\cite{sta99e}.
  The dotted line separates liquid state points from phase separated
  state points, but is {\it not} an indication of the exact $\rho_{\rm
    sp}(T)$, which varies slightly with $T$. (b) Full $\rho$ dependence
  of $D$, also showing the maxima.

Figure 2: Dependence of the rotational relaxations times $\tau_{1}-\tau_{6}$
along $T=240K$ isotherm as a function of density. 

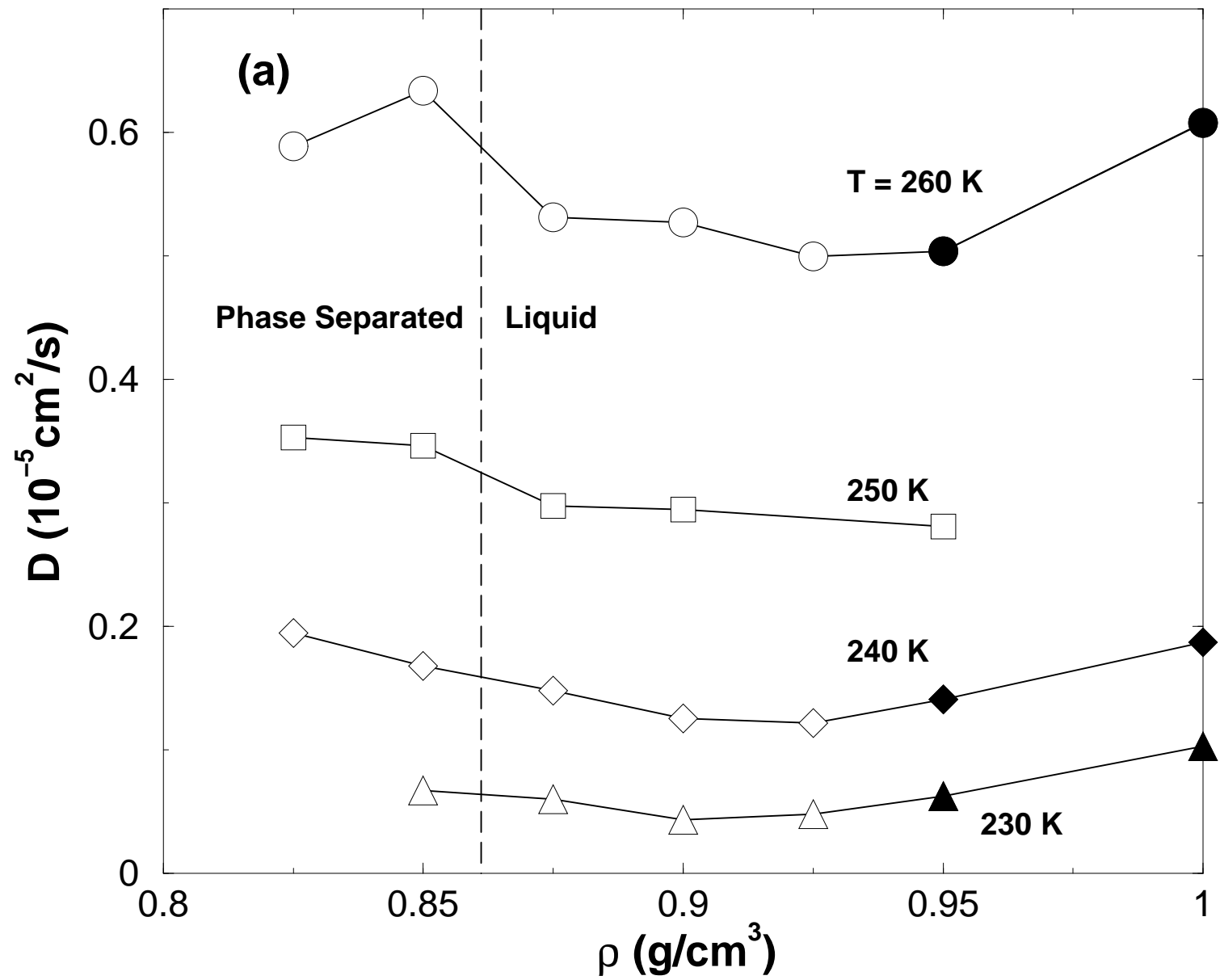
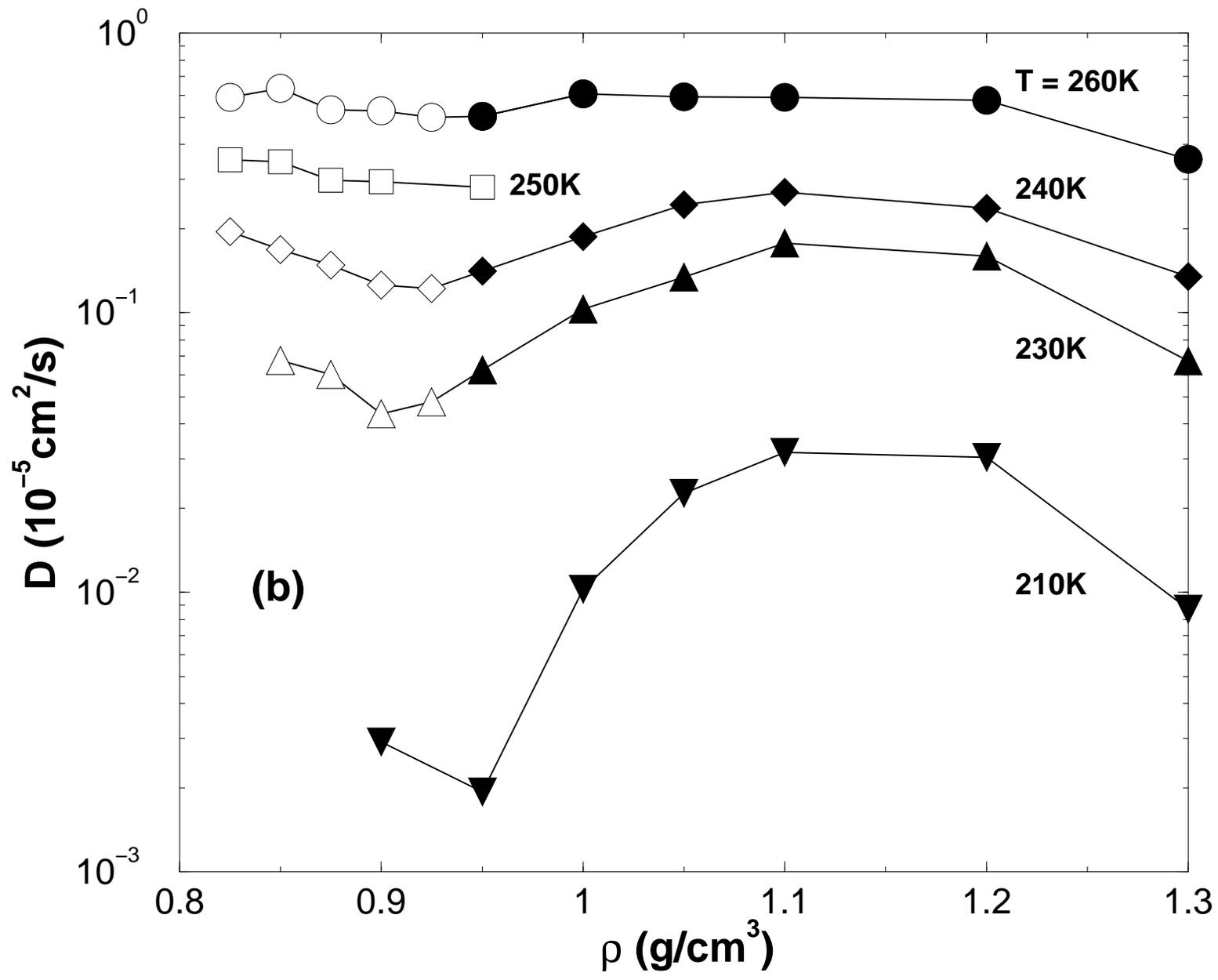
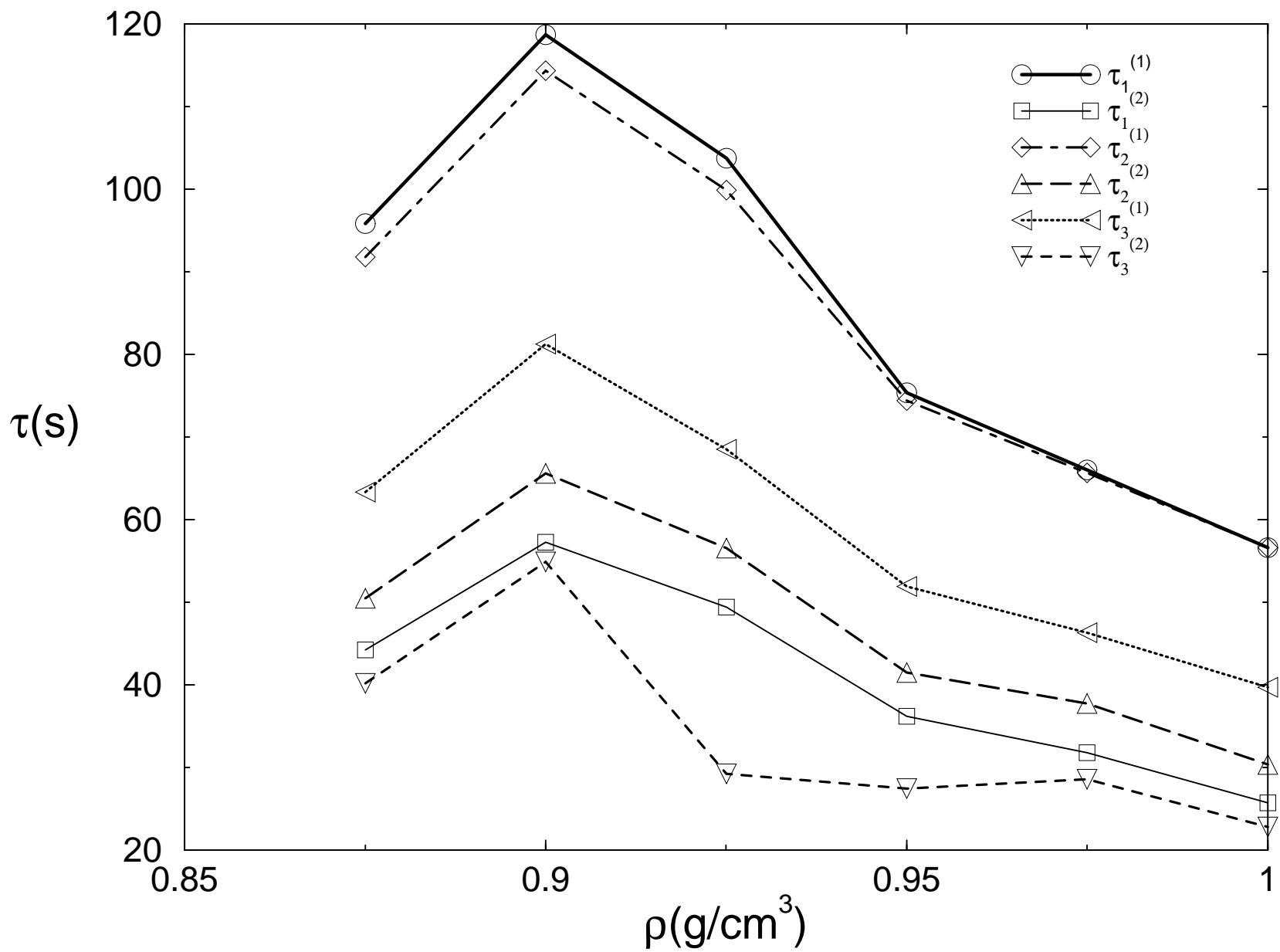
\end{document}